\documentclass[12pt]{iopart}
\usepackage{amsfonts}
\begin{document}
\title{On common solutions of Mathisson equations
under different conditions}
\author{Roman  Plyatsko, Oleksandr Stefanyshyn}
\address{ Pidstryhach Institute of Applied Problems in Mechanics and
Mathematics\\ Ukrainian National Academy of Sciences, 3-b Naukova
Str.,\\ Lviv, 79060, Ukraine}

\ead{plyatsko@lms.lviv.ua}

\begin{abstract}

In the context of investigations of possible highly relativistic
motions of a spinning particle in the gravitational field, which
can be described by the Mathisson equations under different
supplementary condition, we analyze the circular orbits in a
Schwarzschild field. The very orbits most clearly demonstrate the
effect of the gravitational spin-orbit interaction on the
particle's motion. It is shown that the Mathisson equations under
the Frenkel-Mathisson and Tulczyjew-Dixon conditions have the
common solutions describing the highly relativistic circular
orbits in the region $r=3M(1+\delta), |\delta|\ll 1$. These orbits
essentially differ from the geodesic circular orbits in the same
region, particularly by the value of the particle's energy on the
corresponding orbits.

\end{abstract}


\maketitle
\section {Introduction}

The investigations of the behavior of a spinning particle in the
gravitational field are based on the analysis of solutions of the
Mathisson equations \cite{1} and the general relativistic Dirac
equation \cite{2}. In the first case we deal with the description
of the world line and trajectory of a macroscopic body with the
inner rotation, and in the second case with the analysis of the
properties of the wave function of a quantum particle. Under
certain circumstances, in the quasi classical approximation, the
Mathisson equations follow from the Dirac equation \cite{3}. In
this context, it is important that new solutions of the Mathisson
equations stimulate the investigation of the corresponding quantum
states \cite{4}.

The Mathisson-Papapetrou equations can be written in the form \cite{1}
\begin{equation}\label{1}
\frac D {ds} \left(mu^\lambda + u_\mu\frac {DS^{\lambda\mu}} {ds}\right)= -\frac {1} {2}
u^\pi S^{\rho\sigma} R^{\lambda}_{\pi\rho\sigma},
\end{equation}
\begin{equation}\label{2}
\frac {DS^{\mu\nu}} {ds} + u^\mu u_\sigma \frac {DS^{\nu\sigma}} {ds} - u^\nu u_\sigma
\frac {DS^{\mu\sigma}} {ds} = 0,
\end{equation}
where $u^\lambda\equiv dx^\lambda/ds$ is the 4-velocity of a
spinning particle, $S^{\mu\nu}$ is the tensor of spin, $m$ and
$D/ds$ are, respectively, the mass and the covariant derivative;
$R^{\lambda}_{\pi\rho\sigma}$ is the Riemann curvature tensor of
the spacetime. (Greek indices run 1, 2, 3, 4 and Latin indices 1,
2, 3.) After \cite{1} equations (\ref{1}), (\ref{2}) were derived
by many authors using different methods \cite{5}. Often these
equations are named as the Mathisson-Papapetrou-Dixon equations.

An essential circumstance which complicates the study of physical
results following from equations (\ref{1}), (\ref{2}) is the
necessity to supplement these equations by the certain condition
for the description of motions of the particle's center of mass.
However, in the relativistic mechanics in contrast to the
Newtonian mechanics, such condition is not determined uniquely.
Two conditions are usually imposed:
\begin{equation}\label{3}
S^{\lambda\nu} u_\nu = 0,
\end{equation}
or
\begin{equation}\label{4}
S^{\lambda\nu} P_\nu = 0,
\end{equation}
where
\begin{equation}\label{5}
P^\nu = mu^\nu + u_\lambda\frac {DS^{\nu\lambda}}{ds}.
\end{equation}
Relations (\ref{3}) and (\ref{4}) are called the Frenkel-Mathisson
\cite{1,7} and Tulczyjew-Dixon \cite{8,9} conditions
correspondingly. (After \cite{10} relation (\ref{3}) is often
called the Pirani condition.)

Generally, the solutions of equations (\ref{1}), (\ref{2}) at
conditions (\ref{3}), (\ref{4}) are different. For example, in the
Minkowski spacetime equations (\ref{1})--(\ref{3}) have both the
solutions which describe the straight worldlines and the solutions
describing the oscillatory (helical) worldlines \cite{11,12}.
Whereas equations (\ref{1}), (\ref{2}), (\ref{4}) do not admit the
oscillatory solutions. (The interpretation of these unusual
solutions was proposed by M\"oller in the terms of the proper and
non-proper centers of mass \cite{13}.) Nevertheless, there are the
common solutions of equations (\ref{1}), (\ref{2}) under
conditions (\ref{3}) and (\ref{4}) which describe the motions of
the proper center of mass. If the gravitational field is present
and is considered in the post-Newtonian approximation, the
solutions of equations (\ref{1}), (\ref{2}) at (\ref{3}) and
(\ref{4})
 are close with
high accuracy \cite{14}. More generally, the similar situation
takes place if the effect of the particle's spin can be described
by the power in spin corrections to the corresponding expressions
for the geodesic motions \cite{15}.

Special analysis must be carry out in the case when the motion of
the proper center of mass cannot be described by the small
corrections in spin to the geodesic motion. Such a conclusion
follows, for example, from [4,16--18] after the analysis of
solutions of equations (\ref{1}), (\ref{2}), (\ref{3}), where it
is shown that for some highly relativistic motions of a spinning
particle relative to a Schwarzschild mass the world line of this
particle can differ essentially from the geodesic world line.
Therefore, the purpose of the present paper is to investigate the
non-trivial common solutions of equations (\ref{1}), (\ref{2})
under conditions (\ref{3}) and (\ref{4}) in a Schwarzschild field.

It is known that the deviation of the spinning particle motion
from the geodesic one in a Schwarzschild field is caused by the
gravitational spin-orbit interaction \cite{19} (the case of the
highly relativistic motions is considered in \cite{4,16}). As a
result, the maximal effect of the particle's spin on its
trajectory takes place for circular or close to circular motions.
Therefore, in the following we shall study such solutions of
equations (\ref{1}), (\ref{2}), (\ref{4}) which describe the
circular orbits of a spinning particle in the Schwarzschild field.
It is important that these solutions can be obtained in the
analytic form.

Different trajectories of a spinning particles following from
equations (\ref{1}), (\ref{2}), (\ref{4}) in the Schwarzschild and
Kerr fields were investigated in [20--24] depending on the
concrete values of the integrals of energy and angular momentum.
Particularly, in \cite{21} the method of the effective potential
was used for the classification of different orbits, including the
chaotic motions. In contrast to these investigations, we shall
consider the circular orbits only, however, without the
restriction by the small in spin perturbations. That is, the
object of our investigations are the strict equations (\ref{1}),
(\ref{2}), (\ref{4}), and their explicit solutions we shall find
without initial using the integrals of energy and angular
momentum. The values of these integrals will be estimated after
obtaining the explicit expressions for the quantities which
determines the energy and angular momentum.

We stress that the necessary condition for a spinning test
particle \cite{19}
\begin{equation}\label{6}
\frac{|S_0|}{mr}\equiv\varepsilon\ll 1
\end{equation}
will be taken into account in our consideration, where $|S_0|$ is
the absolute value of spin, $r$ is the radial coordinate.

This paper is organized as follows. In Section 2 the general
relations following from equations (\ref{1}), (\ref{2}), (\ref{4})
for any metric are written. The main calculations concerning the
analysis of possible solutions describing the circular orbits in a
Schwarzschild field are presented in Section 3. The region of
existence of the highly relativistic circular orbits and the
particle's energy on these orbits are studied in Sections 4 and 5.
We conclude in Section 6.

\section{Basic relations following from Mathisson equations\\ under condition (\ref{4})}

Usually equations (\ref{1}), (\ref{2}) at condition (\ref{4}) are
written as
\begin{equation}\label{7}
\frac {DP^\lambda}{ds} = -\frac {1} {2} u^\pi S^{\rho\sigma}
R^{\lambda}_{\pi\rho\sigma},
\end{equation}
\begin{equation}\label{8}
\frac {DS^{\lambda\nu}}{ds}=P^\lambda u^\nu - P^\nu u^\lambda.
\end{equation}
The quantity
\begin{equation}\label{9}
\mu=\sqrt{P_\lambda P^\lambda}
\end{equation}
is the mass of a particle, and according to (\ref{7}), (\ref{8})
at condition (\ref{4}), $\mu$ is the integral of motion:
\begin{equation}\label{10}
\frac{d\mu}{ds}=0.
\end{equation}
(Under condition (\ref{3}) the constant quantity is $m$ in
equation (\ref{1})).

There is connection between the 4-velocity $u^\lambda$ and the
4-momentum $P^\lambda$ \cite{21}
\begin{equation}\label{11}
    u^{\lambda}=N\left[\frac{P^{\lambda}}{\mu}+\frac{1}{2\mu^3\Delta}
    S^{\lambda\nu}P^{\pi}R_{\nu\pi\rho\sigma}S^{\rho\sigma}\right],
\end{equation}
where
\begin{equation}\label{12}
\Delta=1+\frac{1}{4\mu^2}R_{\lambda\pi\rho\sigma}S^{\lambda\pi}S^{\rho\sigma},
\end{equation}
\begin{equation}\label{13}
N=\left[1+\frac{1}{4\Delta^2\mu^4}S_{\pi\nu}P_{\lambda}S_{\rho\sigma}R^{\nu\lambda\rho\sigma}
S^{\pi\alpha}P^{\beta}S^{\gamma\delta}R_{\alpha\beta\gamma\delta}
\right]^{-1/2}.
\end{equation}
It follows from (\ref{11}) that $P_\lambda P^\lambda=\mu^2$. The
integral of motion is the value of spin $S_0$:
\begin{equation}\label{14}
S_0^2\equiv \frac{1}{2}S_{\lambda\nu}S^{\lambda\nu}.
\end{equation}

\section{Equations following from (\ref{7}), (\ref{8}) for equatorial circular orbits in a
Schwarzschild field }

We use the Schwarzschild metric in the standard coordinates
$x^1=r,\quad x^2=\theta,\quad x^3=\varphi,\quad x^4=t$ with the
non-zero components of the metric tensor $g_{\lambda\nu}$:
\begin{equation}\label{15}
g_{11}=-\left(1-\frac{2M}{r}\right)^{-1},\quad g_{22}=-r^2,\quad
g_{33}=-r^2\sin^2\theta, \quad g_{44}=1-\frac{2M}{r}.
\end{equation}
Let us consider the relations following from equations (\ref{7}),
(\ref{8}) under condition (\ref{4}) for the equatorial circular
motions $\theta=\pi/2$ of a spinning particle with the constant
angular velocity around the Schwarzschild mass. That is, in
(\ref{7}), (\ref{8}) we put
\begin{equation}\label{16}
u^1=0, \quad u^2=0,\quad u^3=const {\ne} 0, \quad u^4=const {\ne}
0,
\end{equation}
\begin{equation}\label{17}
S^{12}=0, \quad S^{23}=0, \quad S^{13}=const {\ne} 0,
\end{equation}
without any {\it a priori} restriction on $P_\lambda$. According
to (\ref{4}), (\ref{16}), (\ref{17}) we have
\begin{equation}\label{18}
S^{14}=-\frac{P_3}{P_4}S^{13},\quad S^{24}=0,\quad
S^{34}=\frac{P_1}{P_4}S^{13}.
\end{equation}
Then for the derivatives $DS^{\lambda\nu}/ds$ we find
\begin{equation}\label{19}
\frac{DS^{12}}{ds}=0,\quad
\frac{DS^{13}}{ds}=-\frac{P_1}{P_4}\Gamma^1_{44}u^4 S^{13}, \quad
\frac{DS^{23}}{ds}=0,
\end{equation}
where $\Gamma^1_{44}$ is the Christoffel symbol calculated by
metric (\ref{15}). After (\ref{16})--(\ref{19}) we obtain from
(\ref{8}) the two non-trivial relations:
\begin{equation}\label{20}
-\frac{P_1}{P_4}\Gamma^1_{44}u^4S^{13}=P^1u^3,
\end{equation}
\begin{equation}\label{21}
P^2u^3=0.
\end{equation}
Because $u^3=d\varphi/ds\ne 0$, by (\ref{21}) it is necessary
\begin{equation}\label{22}
P^2=0.
\end{equation}
Relation (\ref{20}) is fulfilled in two cases:
\begin{equation}\label{23}
P^1=0,
\end{equation}
or
\begin{equation}\label{24}
-g_{11}u^4\Gamma^1_{44}S^{13}=P_4u^3.
\end{equation}
First we shall analyze  case (\ref{23}). Then the two first
relations (\ref{11}), with $\lambda=1, 2$, are satisfied
identically, and others take the form
\begin{equation}\label{25}
    u^3=N\frac{P^3}{\mu}\left(1-\frac{3S_0^2}{\mu^4\Delta}P_4P^4R^{13}_{\,\,\,\,\, 13}\right),
\end{equation}
\begin{equation}\label{26}
    u^4=N\frac{P^4}{\mu}\left(1+\frac{3S_0^2}{\mu^4\Delta}P_3P^3R^{13}_{\,\,\,\,\, 13}\right),
\end{equation}
where
\begin{equation}\label{27}
\Delta=1+\frac{S^2_0}{\mu^2}R^{13}_{\,\,\,\,\,
13}\left(1-3P_3P^3\right),
\end{equation}
\begin{equation}\label{28}
    N=\left(1+\frac{A}{4\Delta^2\mu^4}\right)^{-1/2},
\end{equation}
\begin{equation}\label{29}
    A\equiv36S^4_0R_{1313}R^{1313}P_3P^3P_4P^4.
\end{equation}
To obtain (\ref{25}), (\ref{26}) we used the expressions
\begin{equation}\label{30}
    \left(S^{13}\right)^2=S^2_0P_4P^4g^{11}g^{33},
    \quad \left(S^{14}\right)^2=S^2_0P_3P^3g^{11}g^{44}
\end{equation}
following from (\ref{4}), (\ref{14}).

Now we shall consider equations (\ref{7}) at conditions
(\ref{16})--(\ref{19}), (\ref{22}), (\ref{23}). The first equation
of set (\ref{7}), with $\lambda=1$, takes the form
\begin{equation}\label{31}
\Gamma^1_{33}u^3P^3+\Gamma^1_{44}u^4P^4=
-g_{44}^2\sqrt{g^{11}g^{33}g^{44}}\frac{S_0}{\mu}\left(2P^3u^4+P^4u^3\right).
\end{equation}
The second equation of (\ref{7}), with $\lambda=2$, is satisfied
identically. From the third and fourth equations of (\ref{7}) it
is easy to obtain correspondingly
\begin{equation}\label{32}
    P^3=const,\quad P^4=const.
\end{equation}

So, all equations (\ref{7}), (\ref{8}) will be satisfied if the
four constant values $u^3, u^4, P^3, P^4$, with relations
(\ref{25}), (\ref{26}), satisfy equation (\ref{31}).

\section{Region of existence of equatorial circular orbits }

From the algebraic relations (\ref{25}), (\ref{26}), (\ref{31}) it
is not difficult to obtain the second order equation for the value
\begin{equation}\label{33}
    x \equiv P_3P^3.
\end{equation}
This equation is:
$$
x^2\left[\left(1-\frac{3M}{r}-\frac{3M^2}{r^2}\varepsilon^2\right)^2-9\varepsilon^2\frac{M^2}{r^2}
\left(1-\frac{2M}{r}\right)\left(1+\frac{4M}{r} \varepsilon^2\right)\right]
$$
$$
+\mu^2x\left[\frac{2M}{r}\left(1-\frac{3M}{r}-\frac{3M^2}
{r^2}\varepsilon^2\right)^2\left(1+\frac{M}{r}\varepsilon^2\right)+9\varepsilon^2\frac{M^2}{r^2}
\left(1-\frac{2M}{r}\right)\left(1+\frac{4M}{r} \varepsilon^2\right)\right]
$$
\begin{equation}\label{34}
+\mu^4\frac{M^2}{r^2}\left(1+\frac{2M}{r}\varepsilon^2\right)=0,
\end{equation}
where
\begin{equation}\label{35}
    \varepsilon^2 \equiv \frac{S_0^2}{\mu^2r^2}
\end{equation}
and according to condition (\ref{6}) it is necessary
$\varepsilon^2\ll 1$. In the trivial case of a spinless particle
($\varepsilon = 0$) from (\ref{34}) we have
\begin{equation}\label{36}
    x=-\mu^2 \frac{M}{r}\left(1-\frac{3M}{r}\right)^{-1}.
\end{equation}
As a result, according to (\ref{33}),
\begin{equation}\label{37}
    P^3=\pm\mu\sqrt{\frac{M}{r^3}}\left(1-\frac{3M}{r}\right)^{-1/2}.
\end{equation}
Equation (\ref{37}) coincides with the known expression for the
4-momentum of a spinless particle with the mass $\mu$ which
follows directly from the geodesic equations for the circular
orbits. Particularly, equation (\ref{37}) shows that such orbits
in a Schwarzschild field exist only at $r>3M$.

It is easy to see that in the case
\begin{equation}\label{38}
    1-\frac{3M}{r}\gg\frac{M^2}{r^2}\varepsilon^2,
\end{equation}
{\it i.e.,} if $r$ is not close to $3M$, it follows from
(\ref{34}) the expression for $P^3$ which differs from (\ref{37})
only by the small correction of order $\varepsilon^2$.

The especial case takes place if $r$ is equal or close to $3M$. Particularly, at
\begin{equation}\label{39}
    r=3M,\quad \epsilon \ne 0
\end{equation}
we obtain from (\ref{34}) the single negative root
\begin{equation}\label{40}
    x=-\frac{1}{\sqrt{3}\mu^2\varepsilon},
\end{equation}
where without loss in generality we put
\begin{equation}\label{41}
    \varepsilon \equiv \frac{S_0}{\mu r},
\end{equation}
{\it i.e.,} $S_0>0$. (According to our choice of the metric
signature, the value $x$ in (\ref{34}) is negative). By
(\ref{33}), (\ref{40}) we have $P^3=\pm
3^{-5/4}\varepsilon^{-1/2}\mu M^{-1}$. However, it is easy to
check that all relations (\ref{25}), (\ref{26}), (\ref{31}) are
satisfied only if
\begin{equation}\label{42}
    P^3=-\frac{\mu}{3^{5/4}M\sqrt{\varepsilon}}.
\end{equation}

Taking into account relations (\ref{25})-(\ref{29}), with the
accuracy to $\varepsilon$ we write $u^3=P^3/\mu$, then according
to (\ref{42}) for the angular velocity we have
\begin{equation}\label{43}
    u^3=- \frac{1}{3^{5/4}M \sqrt{\varepsilon}}.
\end{equation}
If $r$ is not equal to $3M$, however is close to this value:
\begin{equation}\label{44}
    1-\frac{3M}{r}=\delta, \quad |\delta|\ll\varepsilon,
\end{equation}
then instead of (\ref{42}), (\ref{43}) we obtain from (\ref{34})
\begin{equation}\label{45}
    P^3=-\frac{\mu}{3^{5/4}M\sqrt{\varepsilon}}\left(1-\frac{\sqrt{3}\delta}{2\varepsilon}\right),
\end{equation}
\begin{equation}\label{46}
    u^3=- \frac{1}{3^{5/4}M \sqrt{\varepsilon}}\left(1-\frac{\sqrt{3}\delta}{2\varepsilon}\right).
\end{equation}
We stress that here $\delta$ may be both positive and negative.
That is, equation (\ref{34}) admit the circular orbits within the
small neighborhood of $r=3M$, if $|\delta|\ll\varepsilon$, both at
$r>3M$ and $r<3M$.

Now we can compare expressions (\ref{43}), (\ref{46}) with the
corresponding expressions for $u^3$ obtained in \cite{17} from
equations (\ref{1}), (\ref{2}) at condition (\ref{3}). It is easy
to see that these expressions coincide (it is necessary to take
into account that in \cite{17} the notation for $\varepsilon$ was
used $\varepsilon\equiv |S_0|/Mm$, in contrast to (\ref{6})). So,
we conclude that the existence  of the circular orbits of a
spinning particle in the small neighborhood of the value $r=3M$ is
the common result of equations (\ref{1}), (\ref{2}) under
conditions (\ref{3}) and (\ref{4}).

Concerning relation (\ref{24}) it is not difficult to check that
in this case equations (\ref{7}), (\ref{8}) are not compatible.
That is, case (\ref{23}) is the single possible one.

\section{Values of energy and orbital momentum }

Let us estimate the energy $E$ and orbital momentum $L$ of a
spinning particle on the above considered circular orbits. The
expressions of these values for the equatorial motions in a
Schwarzschild field are \cite{25}
\begin{equation}\label{47}
E=P_4+\frac12 g_{44,1}S^{14},
\end{equation}
\begin{equation}\label{48}
L=-P_3-\frac12 g_{33,1}S^{13}.
\end{equation}
According to relations (\ref{6}), (\ref{9}), (\ref{30}),
(\ref{45}) we have
\begin{equation}\label{49}
|g_{44,1}S^{14}|\ll P_4,
\end{equation}
therefore approximately we write
\begin{equation}\label{50}
E=\mu\frac{3^{-3/4}}{\sqrt{\varepsilon}}\left(1-\frac{\sqrt{3}}{2}\frac{\delta}{\varepsilon}\right).
\end{equation}
Let us compare this value of the energy of a spinning particle on
the circular orbit from the small neighborhood of $r=3M$ with the
value of energy of a spinless particle on the geodesic circular
orbit of $r=3M(1+\delta), 0<\delta\ll 1.$ By (\ref{37}) we have
\begin{equation}\label{51}
E_{geod}=\frac{\mu}{3\sqrt{\delta}}.
\end{equation}
It follows from (\ref{50}), (\ref{51}) that
\begin{equation}\label{52}
E^2/\mu^2\gg 1, \quad E^2_{geod}/\mu^2\gg 1,
\end{equation}
{\it i.e.,} these values are highly relativistic. At the same
time, according to (\ref{44}) we have
\begin{equation}\label{52}
E^2/E^2_{geod}= \frac{\sqrt{3}\delta}{\varepsilon}\ll 1,
\end{equation}
that is the energies of the spinning and spinless particles on the
circular orbits from the small neighborhood of $r=3M$ differ to a
great extent. It is not strange because different physical reasons
cause the existence of these orbits: the geodesic circular orbits
exist due to the gravitational attraction, whereas the circular
orbits of a spinning particle are determined by the common action
of the gravitational attraction and the gravitational spin-orbit
repulsion. In addition, we stress that from the point of view of
the comoving observer the acceleration of a spinning particle
relative to a spinless particle is of order $M/r^2$, {\it i.e.,}
is considerable \cite{4}. Therefore, the circular orbits of a
spinning particle from the small neighborhood of $r=3M$ are
essentially non-geodesic orbits.

It is not difficult to check that the similar conclusions follow
from the analysis of the orbital momentum (\ref{48}).

\section{Conclusions}

So, Mathisson equations (\ref{1}), (\ref{2}) at Tulczyjew-Dixon
conditions (\ref{4}) have the solutions which describe highly
relativistic circular orbits of a spinning particle in the small
neighborhood of the value $r=3M$ in a Schwarzschild field, both
for $r>3M$ and $r<3M$. The same orbits are allowed by the
Mathisson equations at Frenkel-Mathisson condition (\ref{3}), that
was shown in \cite{17} while investigating possible synchrotron
radiation of a charged spinning particle in a Schwarzschild field.
It is important that the dynamics of a spinning particle on these
orbits essentially differ from the dynamics of a spinless particle
on the geodesic circular highly relativistic orbits which exist in
the small neighborhood of $r=3M$, at $r>3M$.

The question arise: do equations (\ref{1}), (\ref{2}) under
condition (\ref{4}) allow other types of orbits which are not
circular and differ considerable from the corresponding geodesic
orbits (naturally, when condition (\ref{6}) is satisfied)? To
answer this question it is necessary to carry out more complex
calculation. We point out that the solutions of equations
(\ref{1}), (\ref{2}), (\ref{3}) in a Schwarzschild field, which
describe the essentially non-geodesic non-circular orbits of the
proper center of mass of a spinning particle, were considered in
\cite{18}.

\vskip 3mm


\end{document}